\documentclass[11pt,a4paper]{article}
\usepackage{slashed}
\usepackage{mathtools}
\usepackage{amsfonts}
\usepackage{amssymb}
\usepackage{epsfig}
\usepackage{rotating}
\usepackage{url}
\usepackage{times}
\usepackage{color}
\usepackage{bm}
\usepackage{xcolor,colortbl}
\usepackage{hyperref}
\usepackage[numbers]{natbib}
\usepackage[alsoload=hep]{siunitx}
\usepackage{enumitem}
\usepackage{fancyhdr}
\usepackage{anyfontsize}
\usepackage{empheq}

\newcommand{\ph}{\phantom}
\newcommand{\nn}{\nonumber}
\newcommand{\eps}{\epsilon}

\newcommand{\myt}{\tau}
\newcommand{\E}{\rotatebox[origin=c]{0}{$E$}}
\newcommand{\Ee}{\rotatebox[origin=c]{180}{$E$}}

\newcommand{\idxI}{\scriptscriptstyle{A}}
\newcommand{\idxJ}{\scriptscriptstyle{B}}
\newcommand{\idxK}{\scriptscriptstyle{C}}
\newcommand{\idxL}{\scriptscriptstyle{D}}
\newcommand{\idxM}{\scriptscriptstyle{E}}
\newcommand{\idxN}{\scriptscriptstyle{F}}

\newcommand{\idxa}{a}
\newcommand{\idxb}{b}
\newcommand{\idxc}{c}

\newcommand{\idxi}{i}
\newcommand{\idxj}{j}
\newcommand{\idxk}{k}

\DeclareMathSizes{10}{10}{7}{6}

\title{Spacetime and dark matter from spontaneous breaking of Lorentz symmetry}

\author{Tom~Z\l o\'{s}nik$^{1}$ \footnote{zlosnik@fzu.cz}, Federico Urban$^{1}$ \footnote{federico.urban@fzu.cz}, Luca Marzola$^{2}$ \footnote{luca.marzola@cern.ch }, and Tomi Koivisto$^{2,3}$\footnote{tomi.koivisto@nordita.org}
\\{\small \it	$^1$ CEICO, Institute of Physics of the Czech Academy of Sciences, Na Slovance 1999/2, 182 21, Prague} 
\\{ \small \it	$^2$ National Institute of Chemical Physics and Biophysics, R\"avala pst.\ 10, 10143 Tallinn, Estonia.}
\\{ \small \it$^3$ Nordita, KTH Royal Institute of Technology and Stockholm University, Roslagstullsbacken 23, 10691 Stockholm, Sweden}}

\begin{document}
\maketitle

\abstract{
	It is shown that a spontaneously-broken gauge theory of the Lorentz group contains Ashtekar's chiral formulation of General Relativity accompanied by dust. From this perspective, gravity is described entirely by a  connection $\omega^{\idxI}_{\ph{\idxI}\idxJ}=\omega^{\idxI}_{{\ph\idxI}\idxJ\mu}dx^{\mu}$ valued in the Lie algebra of the complexified Lorentz group and a Lorentz-charged scalar field $\myt^{\idxI}$. The model is ``pre-geometric'' in the sense that the spacetime metric may be constructed only in the symmetry-broken regime. We speculate on the extent to which this dust may account for dark matter and on the behaviour of the theory in regimes where the symmetry is not broken.}

\section{Introduction and results}

Lorentz symmetry is a cornerstone of modern physics. The Lagrangians of particle physics are invariant under global Lorentz transformations. When matter fields are coupled to gravity in the Einstein-Cartan-Sciama-Kibble (ECSK) formulation~\cite{Kibble1961}, this is promoted to a local Lorentz invariance. Nonetheless, Lorentz invariance is broken in many physical systems. The possibility that additional fields may exist in the gravitational sector that spontaneously break Lorentz symmetry has attracted a lot of attention recently, for example in the Einstein-Aether~\cite{Jacobson2000} and ghost condensate~\cite{ArkaniHamed2003} models.

As a separate issue, the extent to which gravity can be formulated in a manner reminiscent of the theories of particle physics has been on ongoing area of research. In what sense is gravity a gauge theory \cite{HehlEtAl1994,Krasnov2011,Celada2015,Krasnov2017}? It was discovered~\cite{MacDowell1977,Stelle1979} that the ECSK theory can arise as a limit of a spontaneously-broken gauge theory whose mathematical ingredients are a gauge field/connection for the de Sitter (or anti-de Sitter) group, and a ``gravitational Higgs'' scalar field in the fundamental representation of the group which breaks the symmetry down to that of $SO(1,3)$, the Lorentz group; here the metric tensor arises as a composite object built from the connection and the symmetry breaking field. These approaches are very closely related to a formulation of geometry due to Cartan~\cite{Wise2006,Westman:2014yca}.

We will look to implement Lorentz symmetry breaking in the gravitational sector by adopting a similar description. However, instead our variables will be taken to be the gauge field/connection $\omega^{\idxI}_{\ph{\idxI}\idxJ} = \omega^{\idxI}_{\ph{\idxI}{\idxJ}\mu}dx^{\mu}$ for the \emph{complexified} Lorentz group $SO(1,3)_{C}$, and a Lorentz-charged scalar field $\myt^{\idxI}$ which is free to break the Lorentz symmetry down to $SO(3)_{C}$ if $\myt_{\idxI}\myt^{\idxI}<0$~\footnote{See the Appendix for notation and some useful formulas.}; notice that we \emph{do not} introduce any additional quantities, such as the co-tetrad soldering form or the metric tensor. Nonetheless, the gravitational effect of the fields will be shown to be that of General Relativity (GR) coupled to an additional dust component in the universe and we will examine the extent to which this may be the origin of effects attributed to dark matter (DM). We consider the following polynomial action~\footnote{The action~(\ref{action}) is $\int d^{4}x \eps^{\mu\nu\alpha\beta}\eps_{\idxI\idxJ\idxK\idxL}D_{\mu}\myt^{\idxI}D_{\nu}\myt^{\idxJ}R^{+\idxK\idxL}_{\ph{+\idxK\idxL}\alpha\beta}/4$ when written in standard notation, where $\eps^{\mu\nu\alpha\beta}$ is the Levi-Civita density and $\eps_{\idxI\idxJ\idxK\idxL}$ is the completely antisymmetric $SO(1,3)$-invariant (as well as $SO(1,3)_{C}$-invariant).}:
\begin{align}
S_{G}[\omega,\myt]& =\int {\cal L}_{G} = i \int  D\myt_{\idxI} D\myt_{\idxJ} R^{+ {\idxI\idxJ}} \label{action}
\end{align}
where
\begin{align}
D\myt^{\idxI} &\equiv d\myt^{\idxI} + \omega^{\idxI}_{\ph{\idxI}\idxJ}\myt^{\idxJ} \\
R^{+\idxI\idxJ} &\equiv d\omega^{+\idxI\idxJ} + \omega^{+\idxI}_{\ph{+\idxI}\idxK}\omega^{+\idxK\idxJ} \,.
\end{align}
Here $\omega^{+\idxI\idxJ}$ is the \emph{self-dual} part of $\omega^{\idxI\idxJ}$,
 $D\myt^{\idxI} = D_{\mu}\myt^{\idxI} dx^{\mu}$, $R^{+\idxI\idxJ} = \frac12 R^{+\idxI\idxJ}_{\ph{+\idxI\idxJ}\mu\nu}dx^{\mu}dx^{\nu}$, and forms are multiplied via the wedge product.

Our results can be summarised in the following statement: the action~(\ref{action}) contains the classical solutions described by the more familiar~\cite{Lim:2010yk,Golovnev:2013jxa,Ramazanov:2016xhp,Jacobson:2014mda,SebastianiEtAl2016}
\begin{align}
S_{G}'[g,\myt,\rho] &= \frac{1}{2}\int \sqrt{-g}d^{4}x\bigg( R - \rho(\partial^{\mu}\myt\partial_{\mu}\myt+1)\bigg) \label{grdust}
\end{align}
where $\rho$ is a Lagrange multiplier, $-\myt^{2}\equiv \myt^{\idxI}\myt_{\idxI}$, and $R\sqrt{-g}$ is the usual Einstein-Hilbert term for the metric $g_{\mu\nu}$, \emph{in a broken symmetry phase} in which $\myt_{\idxI}\myt^{\idxI}<0$\footnote{No gauge-fixing or splitting is needed to obtain this result.}.  Once we arrive at this equation (and given assumptions the coupling of gravitational fields to matter that we will later discuss) the cosmology and phenomenology of this model is in every aspect identical to that of GR plus mimetic DM.

\section{Hamiltonian formulation}
\label{sec:hform}

To make contact with more familiar models we will perform a $3+1$ decomposition of fields and construct the Hamiltonian for the theory. We assume that the manifold $M$ is topologically $R\times\Sigma$ for some submanifold $\Sigma$: these surfaces correspond to $t=\mathrm{cst.}$ where $t(x^{\mu})$ is a global time function. We may additionally define a `flow of time' vector $t^{\mu}$ which satisfies $t^{\mu}\partial_{\mu}t = 1$. As such we can decompose the fields $\{\myt^{\idxI},\omega^{\idxI\idxJ}\}$ and their exterior derivatives as:
\begin{align}
\omega^{\idxI\idxJ} &= \Omega^{\idxI\idxJ} dt + \bar{\omega}^{\idxI\idxJ}\\
d\myt^{\idxI} &= \partial_{t}\myt^{\idxI}dt + \bar{d}\myt^{\idxI}\\
d\omega^{\idxI\idxJ} &= \partial_{t}\bar{\omega}^{\idxI\idxJ}_{\ph{\idxI\idxJ}\idxa} dtdx^{\idxa} + \bar{d}\Omega^{\idxI\idxJ}dt+\bar{d}\bar{\omega}^{\idxI\idxJ}
\end{align}
where $t^{\mu}\bar{\omega}^{\idxI\idxJ}_{\ph{\idxI\idxJ}\mu} =0$ and $\bar{d}$ is the exterior derivative according to the spatial coordinates $x^{\idxa}$. The Lagrangian ${\cal L}_{G}$ four-form from~(\ref{action}) can then be shown to be:
\begin{align}
{\cal L}_{G} \overset{b}{=} &\, i dt \bigg( \bar{D}\myt^{\idxI}\bar{D}\myt^{\idxJ}\partial_{t}\bar{\omega}^{+}_{{\idxI\idxJ}\idxa} dx^{\idxa}+ 2\bar{D}\myt^{\idxJ}\bar{R}^{+}_{\idxI\idxJ}\partial_{t}\myt^{\idxI} \bigg) \nn\\
+ &\, i dt \bigg( 2\Omega^{\idxI\idxK}\myt_{\idxK}\bar{D}\myt^{\idxJ}\bar{R}^{+}_{\idxI\idxJ} + \Omega^{+}_{\idxI\idxJ}\bar{D} \big(\bar{D}\myt^{\idxI}\bar{D}\myt^{\idxJ}\big) \bigg) \label{lag1}
\end{align}
where $\overset{b}{=}$ means equal to up to a boundary term and $\bar{D}$ is the pullback to $\Sigma$ of the covariant derivative according to $\bar{\omega}^{\idxI\idxJ}$ so that $\bar{D} \equiv \bar{d} + \bar{\omega}$. To make further progress we employ a partial gauge fixing:
\begin{align}
\bar{D}\myt^{0} = \bar{d}\myt^{0}+\bar{\omega}^{0\idxi}\myt_{\idxi}\overset{*}{=} 0  \label{timegauge}
\end{align}
where $\overset{*}{=}$ means an equation satisfied in this particular gauge. The gauge condition~(\ref{timegauge}) is analogous to the ``time gauge'' employed in Ashtekar's theory of gravity and similarly it will serve to simplify the treatment of the kinetic term for the connection~\cite{Romano1991}; we will remain in this gauge as follows.

To simplify expressions we define the new fields $\{\E^{\idxi},\pi_{0},\pi_{\idxi}\}$: 
\begin{align}
\E^{\idxi} &\equiv \bar{D}\myt^{\idxi}\\
\pi_{0} &\equiv \eps_{0\idxj{\idxK\idxL}}\E^{\idxj} \bar{R}^{+\idxK\idxL} \\
\pi_{\idxi} &\equiv \eps_{\idxi\idxj{\idxK\idxL}}\E^{\idxj} \bar{R}^{+\idxK\idxL} 
\end{align}
where we've used the identity $iR^{+\idxI\idxJ} = \frac12\eps^{\idxI\idxJ}_{\ph{\idxI\idxJ}\idxK\idxL}R^{+\idxK\idxL}$. In what follows, these definitions are implemented by the use of Lagrangian constraints with associated Lagrange multiplier fields $C_{\idxi},Ndt,N^{\idxi}dt$, respectively. We will see that the field $\E^{\idxi}$ will play the role of the spatial co-triad whilst $\pi_{0}$ and $\pi_{\idxi}$ appear in the combination $dt \pi_{0} \partial_{t}\myt^{0}+dt \pi_{\idxi}\partial_{t}\myt^{\idxi} $ and so can be interpreted as canonical momenta of $\{\myt^{0},\myt^{\idxi}\}$. The fields $N$ and $N^{\idxi}$ will play the role of the lapse and shift~\cite{Arnowitt1962} of the spacetime metric that emerges in this model. 

Now, $\{\Omega^{+\idxI\idxJ},\Omega^{-\idxI\idxJ}\}$ are independent fields~\footnote{If $\omega^{\idxI\idxJ}$ were the connection of the real Lorentz group $SO(1,3)$ then we would to have additionally ensure that $\omega^{-\idxI\idxJ} = \omega^{*+\idxI\idxJ}$ so that $\omega^{\idxI\idxJ}$ would be real.}. The Lagrangian is linear in $\{\Omega^{+\idxI\idxJ},\Omega^{-\idxI\idxJ}\}$: from the equations of motion obtained from variations with respect to $\Omega^{-\idxI\idxJ}$ and the constraints enforced by $\{N,N^{\idxi}\}$ we recover the following equations:
\begin{align}
\pi_{0} \myt_{\idxi} - \pi_{\idxi}\myt_{0} &=0 \label{pit1}\\
\pi_{[\idxi}\myt_{\idxj]} &=0  \label{pit2}
\end{align}
where we have assumed that $\myt^{\idxI}$ and its conjugate $\pi^{\idxI}$ are real~\footnote{In principle we could keep these variables complex and impose the reality conditions only at the very end; however this would not change our results at the expense of a hazier physical formulation.}.

Equation~(\ref{pit2}) is automatically satisfied if~(\ref{pit1}) is used and we may then use~(\ref{pit1}) to eliminate $\pi_{\idxi}$ from the Lagrangian. Furthermore, we will look to use the field $\myt$ instead of $\myt^{0}$; they are related via the definition of $\myt$: $\myt^{0}= \pm \sqrt{\myt^{2}+\myt^{\idxi}\myt_{\idxi}}$. We choose the positive square root from now on; this implies additionally that $\myt>0$, interpreted as the `time' component of $\myt^{\idxI}$ in the gauge $\myt^{\idxI}\overset{\bullet}{=} (\myt,0,0,0)$.
Given these results, it can be seen that $\E_{\idxi}\myt^{\idxi} = -\myt \bar{d}\myt$; hence if $\E_{\idxi\idxa}$ is invertible in the sense that there exists a matrix $\Ee^{\idxb\idxi}$ such that $\E_{\idxi\idxa}\Ee^{\idxa\idxj} = \delta^{\idxj}_{\idxi}$ then we can solve this equation to recover:
\begin{align}
\myt^{\idxi} &= -\myt \Ee^{\idxi}\cdot\bar{d}\myt \label{taui} \,.
\end{align}
The kinetic term of $\myt^{\idxI}$ can then be simplified as $dt \pi_{\idxi}\partial_{t}\myt^{\idxi} + dt \pi_{0} \partial_{t}\myt^{0} = dt \pi_{0}\frac{\myt}{\myt_{0}}\partial_{t}\myt\equiv dt{\cal P}\partial_{t}\myt$, where ${\cal P}$ is the canonical momentum three-form for the field $\tau$. Together with the constraints~(\ref{pit1}) and~(\ref{pit2}), the Lagrangian simplifies to
\begin{align}
{\cal L}_{G} &\overset{b}{=} \frac12 dt\eps_{\idxi\idxj\idxK\idxL} \E^{\idxi}\E^{\idxj}\partial_{t}\bar{\omega}^{+\idxK\idxL}_{\ph{+\idxK\idxL}\idxa} dx^{\idxa}+dt{\cal P}\partial_{t}\myt \nn\\
&+dt \Omega^{0\idxi +} \eps_{0\idxi\idxj\idxk}\bar{D}^{+}\big(\E^{\idxj}\E^{\idxk} \big) +C_{\idxi}\left(\E^{\idxi}-\bar{D}\myt^{\idxi}\right)   \nn\\
&+Ndt(\pi_{0} - \eps_{0\idxj\idxK\idxL}\E^{\idxj} \bar{R}^{+\idxK\idxL})  +N^{\idxi}dt(\pi_{\idxi} - \eps_{\idxi\idxj\idxK\idxL} \E^{\idxj} \bar{R}^{+\idxK\idxL}) \,. \label{lag3}
\end{align}
Varying with respect to the anti self-dual part of $\bar\omega^{\idxI\idxJ}_{\ph{\idxI\idxJ}\idxa}$ we find that $C_{\idxi}=0$; finally, using the results~(\ref{taui}) and~(\ref{pit1}) we conclude that:
\begin{align}
S_{G} &\overset{b}{=} \int dt\bigg(\eps_{\idxi\idxj\idxk} \E^{\idxi}\E^{\idxj}\partial_{t}\bar{\omega}^{+0\idxk}_{\ph{+0\idxk}\idxa} dx^{\idxa}+{\cal P}\partial_{t}\myt - {\cal H}\bigg) \label{lag4}
\end{align}
where the Hamiltonian three-form ${\cal H}$ is given by:
\begin{align}
{\cal H }&= \Omega^{+\idxi0} \eps_{\idxi\idxj\idxk}\bar{D}^{+}\big(\E^{\idxj}\E^{\idxk} \big)  +N({\cal P}\sqrt{1+\partial^{\idxa}\myt\partial_{\idxa}\myt}+ \eps_{\idxi\idxj\idxk}\E^{\idxi} \bar{R}^{+\idxj\idxk})\nn\\
&+N^{\idxi}({\cal P}\partial_{\idxi}\myt +2 \eps_{\idxi\idxj\idxk} \E^{\idxj} \bar{R}^{+0\idxk} ) \label{ham1}
\end{align}
where $\eps_{\idxi\idxj\idxk} \equiv \eps_{0\idxi\idxj\idxk}$, $\partial_{\idxi}\myt \equiv \Ee^{\idxa}_{\ph{\idxa}\idxi}\partial_{\idxa}\myt$, $\partial^{\idxa}\myt\equiv h^{\idxa\idxb}\partial_{\idxb}\myt$, $h^{\idxa\idxb}\equiv \Ee^{\idxa}_{\ph{\idxa}\idxi}\Ee^{\idxb\idxi}$. The action~(\ref{lag4}) corresponds to the Hamiltonian formulation of the self-dual covariant formulation of Ashtekar's theory of gravity~\cite{Ashtekar1986} coupled to a rotationless dust fluid~\cite{Brown1994,GieselThiemann2012}. As we have anticipated, the equations of motion resulting from~(\ref{lag4}) have solutions that include those that follow from the action~(\ref{grdust}).

The action~(\ref{lag4}) has 9 complex variables in $\omega^{+\idxi0}$ and further 9 complex ones in $\E^{\idxj}$, plus the real $\myt$ and ${\cal P}$. By inspection, we see that there are Lagrange multipliers for $\Omega^{+\idxI\idxJ}$ (3 complex constraints) and $(N,N_{\idxi})$ (4 complex constraints), and we have equations~(\ref{pit1}) and~(\ref{taui}) which combine to eliminate 6 more phase space variables. Hence, following standard Dirac counting, the propagating degrees of freedom are two complex and one real, which, by means of additional reality conditions we can reduce to the two real degrees of freedom of GR plus one for rotationless dust. Notice that the reality conditions can be consistently imposed within this framework~\cite{MoralesTecotl:1996sh,Romano:1991up}.

An alternative statement of the path to this conclusion is as follows: we can proceed starting from the original action~(\ref{action}) where there are 24 complex variables in $\omega^{\idxI\idxJ}_{\ph{\idxI\idxJ}\mu}$, but since the $\omega^{-\idxI\idxJ}$ do not appear in $R^{+\idxI\idxJ}$ they will not appear with time derivatives acting on them; moreover, in the $3+1$ decomposition we see that $\Omega^{+\idxI\idxJ}$ do not have time derivatives acting on them, so they will drop out as well leaving us with 9 complex Lagrangian connection variables (which correspond to 18 complex phase space variables); on top of that there are 4 complex variables in $\myt^{\idxI}$ which correspond to 8 complex phase space variables, for a total or 26 (complex) ones. The constraints in this case (without imposing reality conditions) come from $\Omega^{+\idxI\idxJ}$ and $\Omega^{-\idxI\idxJ}$ (6 complex constraints), and $(N,N_{\idxi})$ (4 complex constraints). If it is the case that all constraints mutually commute with one another on the constraint surface they define~\cite{Romano:1991up}, the final result is 3 complex propagating degrees of freedom, which we can reduce to the usual 3 via reality conditions.

We additionally note that it is possible to reformulate all our findings in terms of two \emph{independent} connections of the \emph{real} $SO(1,3)$ Lorentz group. In this case the symmetry-breaking scalar $\myt^{\idxI}$ is real and the reality conditions are not needed. This formulation however lacks the clarity of the complexified Lorentz group case. Lastly, in the chiral formulation with the complex Lorentz group, the (partial) chirality of the original action, where $\omega^{+\idxI\idxJ}$ and $\omega^{-\idxI\idxJ}$ do not appear symmetrically in the action, is not present in~(\ref{grdust}), so, for instance, gravity waves of different chirality propagate as in GR, and both polarisations are present.

\section{As preferred frame}
\label{sec:frame}

Systems of GR coupled to a rotationless dust as in~(\ref{grdust}) define a preferred frame corresponding to families of observers whose four velocity $U^{\mu}$ is $\partial^{\mu}\myt$~\cite{Brown1994}. In this frame, $\myt$ tells the time. If again we assume $\myt_{\idxI}\myt^{\idxI}<0$ and that we may choose $t=\myt \equiv \sqrt{-\myt_{\idxI}\myt^{\idxI}}$ then from~(\ref{taui}) we see that $\myt^{\idxi}=0$ and hence the gauge~(\ref{timegauge}) and that defined by the condition $\myt^{\idxI} \overset{\bullet}{=} (\myt,0,0,0)$ coincide and it may be shown that the pullback $\bar{\omega}^{\idxI\idxJ}$ of $\omega^{\idxI\idxJ}$ to surfaces of constant $\myt$ for the choice $\gamma = i$ takes the form:
\begin{align}
\bar{\omega}^{\idxI\idxJ} \overset{*}{=} \left( \begin{array}{cc}
0 & \frac{1}{\myt}\E^{\idxi}\\
- \frac{1}{\myt}\E^{\idxi}\ & \eps^{\idxi\idxj\idxk}\bigg(\Gamma_{\idxk}- i\left(K_{\idxk}- \frac{1}{\myt}\E_{\idxk}\right)\bigg)
\end{array} \right)
\label{barom}
\end{align}
where $E^{\idxi}\equiv D\myt^{\idxi}$, $\eps^{\idxi\idxj\idxk}\Gamma_{\idxk}(\E,\bar{\partial}\E)$ is the torsion-free $SO(3)_{C}$ connection compatible with $\E^{\idxi}$ and $K_{\idxi}$ is the extrinsic curvature form. We may identify a spatial metric $h_{\idxa\idxb} \equiv \delta_{\idxi\idxj}\E^{\idxi}_{\ph{\idxi}\idxa}\E^{\idxj}_{\ph{\idxj}\idxb}$ and extrinsic curvature tensor $K_{\idxa\idxb} =\frac12{\cal L}_{(n^{\mu})} h_{\idxa\idxb}$ where $n^{\mu}$ is the unit normal to surfaces of constant $\myt$. The extrinsic curvature form $K_{\idxi}$ is related to $K_{\idxa\idxb}$ via $K_{\idxi} = K_{\idxa\idxb}\Ee^{\idxb}_{\ph{\idxb}\idxi}dx^{\idxa}$. If we assume that $\E^{\idxi}$ is real then in the preferred frame the real part of $\bar{\omega}^{0\idxi}$ tells us about distances and angles on the hypersurface $\Sigma$ whilst the real part of $\bar{\omega}^{\idxi\idxj}$ contains information about parallel transport of $SO(3)_{C}$ vectors on $\Sigma$ and the intrinsic curvature of the surface. Information about evolution of the spatial metric with respect to the preferred time is contained in the imaginary part of $\bar{\omega}^{\idxi\idxj}$: this field is related to how much the extrinsic curvature $K_{\idxa\idxb}$ differs from $h_{\idxa\idxb}/\myt$.
 
By way of comparison, in the ECSK theory, where the gravitational field is described by a frame field $e^{\idxI}=e^{\idxI}_{\ph{\idxI}\mu}dx^{\mu}$ alongside the $SO(1,3)$ (real) connection $\omega^{\idxI\idxJ}$, we have in the pure gravity case in the gauge $e^{0}_{\idxa}\overset{\star}{=}0$ that:
\begin{align}
 \bar{\omega}_{(ECSK)}^{\idxI\idxJ} \overset{\star}{=} \left( \begin{array}{cc}
 0 & K^{\idxi}\\
 - K^{\idxi} & \eps^{\idxi\idxj\idxk}\Gamma_{\idxk}
 \end{array} \right)  , \quad
 \bar{e}_{(ECSK)}^{\idxI} \overset{\star}{=} \left(\begin{array}{c} 0 \\ \E^{\idxi} \end{array} \right) \,.
\end{align}
We note that though the forms of $\bar{\omega}^{\idxI\idxJ}$ and $\bar{\omega}^{\idxI\idxJ}_{(ECSK)}$ are very different, nonetheless $\bar{\omega}^{+\idxI\idxJ} = \bar{\omega}^{+\idxI\idxJ}_{(ECSK)}$ in these gauges. Furthermore, the form~(\ref{barom}) has much in common with the proposed Cartan connection for the formalism presented in~\cite{GielenWise2011} where a `non-dynamical' $\myt^{\idxI}$ field was introduced alongside fields $\{e^{\idxI},\omega^{\idxI\idxJ}\}$ to enable a Cartan-geometric interpretation of geometrodynamics.

The action~(\ref{action}) is a specific case (when $\gamma = i$) of the following action
\begin{align}
S_{G}[\omega,\myt] &= \frac12\int \left(\frac12\eps_{\idxI\idxJ\idxK\idxL} + \frac{1}{\gamma}\eta_{\idxI\idxK}\eta_{\idxJ\idxL}\right) D\myt^{\idxI}D\myt^{\idxJ}R^{\idxK\idxL} \label{gammaaction} \\
&\overset{b}{=}\frac12\int \left(\frac18 \myt^2 \eps_{\idxI\idxJ\idxK\idxL} - \frac{1}{\gamma}\myt_{\idxI}\myt_{\idxK}\eta_{\idxJ\idxL}\right) R^{\idxI\idxJ}R^{\idxK\idxL}
\end{align}
which is equivalent to the Ashtekar-Barbero-Holst action~\cite{Holst1995} under the replacement $D\myt^{\idxI}\rightarrow e^{\idxI}$.

In the frame where $t=\myt$, it can be shown that the action~(\ref{gammaaction}) produces equations of motion for the spatial metric $h_{\idxa\idxb}$ that would follow from the following Lagrangian:
\begin{align}
L &=  \sqrt{h} \bigg(R^{(3)}+\frac{1}{\gamma^{2}}\left(K^{2}-K^{\idxa\idxb}K_{\idxa\idxb}\right)\bigg) \label{ga2}
\end{align}
where $R^{(3)}$ is the Ricci scalar built from the Christoffel symbols according to $h_{\idxa\idxb}$. This illustrates why it is only when $\gamma^{2} = -1$ that GR is recovered as part of the model, as it is only in this case that the ADM Lagrangian is recovered (with lapse function $N=1$)~\cite{Arnowitt1962}.  From~(\ref{ga2}) it is obvious that $\gamma=-i$ would give identical results.

By contrast, in the limit where $\gamma\rightarrow \infty$, the Lagrangian’s dependence on extrinsic curvature disappears and hence the metric does not have `dynamics' in the usual sense~\footnote{The authors of~\cite{Aldrovandi2004} consider the action~(\ref{action}) in the limit $\gamma\rightarrow \infty$ alongside a term $\eps_{\idxI\idxJ\idxK\idxL}D\myt^{\idxI}D\myt^{\idxJ}D\myt^{\idxK}D\myt^{\idxL}$. We do not find agreement with their conclusion that this combined action yields GR in the presence of a cosmological constant.}.

\section{As Dark Matter}
\label{sec:dm}

The Lagrangian~(\ref{grdust}) corresponds to that of an interesting, recent candidate to explain the effects attributed to DM, often referred to as \emph{mimetic DM}~\cite{Chamseddine:2013kea,Chamseddine2014,SebastianiEtAl2016}, where $-\partial^{\mu}\myt$ is the four-velocity of the DM and $\rho$ its density. The `darkness' of the dust in~(\ref{grdust}) depends on the manner in which the gravitational fields $\{\myt^{\idxI},\omega^{\idxI\idxJ}\}$ couple to matter fields. Firstly, it can be shown that the effect of adding the following action
\begin{align}
S_{\Lambda} &=  -\frac{\Lambda}{24} \int \eps_{\idxI\idxJ\idxK\idxL}D\myt^{\idxI}D\myt^{\idxJ}D\myt^{\idxK}D\myt^{\idxL}
\end{align}
to~(\ref{action}) is to result in a cosmological constant contribution to~(\ref{grdust}). Note that here, the familiar action for a cosmological constant is recovered under the replacement $D\myt^{\idxI} \rightarrow e^{\idxI}$.

Indeed, we expect that the recovery of familiar matter couplings to gravity would be achieved in the present case by taking the matter actions considered~\cite{Ashtekar1989} in the context of the Ashtekar formulation of gravity and everywhere replacing $e^{\idxI} \rightarrow D\myt^{\idxI}$. If so, the gravitational effect of $\myt$ will remain that of dust not coupled directly to other matter.

It has been argued that unless the mimetic DM action is modified by terms breaking the action's shift symmetry $\myt\rightarrow \myt + \mathrm{cst.}$ in~(\ref{grdust})~\cite{Mirzagholi2014} then the DM abundance we observe is not consistent with common models of inflation.  In our model~(\ref{action}), it may be necessary then to consider couplings between $\myt^{\idxI}$ and matter which break the shift symmetry of~(\ref{grdust}). Another open and important question for this model is the issue of caustics, which are typically formed in certain models of DM~\cite{Babichev2016,Babichev2017}. We note that the action~(\ref{action}) is more general than the theory~(\ref{grdust}) in that other phases can exist (i.e. those with $\myt_{\idxI}\myt^{\idxI}\geq 0$) which will display different behaviour; it is however unclear whether classical evolution can connect different phases and, if so, whether this would affect caustic formation.

\section{Discussion}
\label{sec:disc}

Although the action~(\ref{action}) results in a familiar theory in a symmetry-broken regime, one may wonder whether there can exist phases where $\myt^{\idxI} = 0$ and whether they can dynamically evolve to regions where the Lorentz symmetry is broken. The classical equations of motion for general values of $\gamma$ are given by:
\begin{align}
0 &=  -P_{\idxI\idxJ[\idxL\idxM]}R^{\idxI}_{\ph{\idxI}\idxN}\myt^{\idxN}D\myt^{\idxJ} + \myt_{[\idxL}P_{\idxM]\idxI\idxJ\idxK} D\myt^{\idxI}R^{\idxJ\idxK} \label{eqmo1}\\
0 &= P_{\idxI\idxJ\idxK\idxL}R^{\idxJ}_{\ph{\idxJ}\idxM}\myt^{\idxM}R^{\idxK\idxL} \label{eqmo2}
\end{align}
where
\begin{align}
P_{\idxI\idxJ\idxK\idxL} &\equiv \frac12\left(\frac12\eps_{\idxI\idxJ\idxK\idxL} + \frac{1}{\gamma}\eta_{\idxI\idxK}\eta_{\idxJ\idxL}\right) \,.
\end{align}
By inspection, there indeed exist solutions to~(\ref{eqmo1}) and~(\ref{eqmo2}) where $\myt^{\idxI}=0$ globally and $\omega^{\idxI\idxJ}$ is unrestricted. Another, `complimentary' class of solutions that exist for ~(\ref{eqmo1}) and~(\ref{eqmo2}) are those for which $R^{\idxI\idxJ}=0$ globally; for such solutions, there exist gauges where $\omega^{\idxI\idxJ} \overset{*}{=} 0$ and $\tau^{\idxI}$ is unrestricted. Among these solutions, those with $\tau^{\idxI}\tau_{\idxI} <0$ globally correspond to a dark matter density $\rho = 0$. Another interesting set of solutions from this class are those for which $D\tau^{\idxI} \overset{*}{=} d\tau^{\idxI}$ is an invertible matrix in indices $\{{\idxI}, \mu \}$ - we may then choose coordinates such that $\partial_{\mu}\tau^{\idxI} = \delta^{\idxI}_{\mu}$ and hence the $\tau^{\idxI}$ take the form of Minkowski coordinates according to the `metric' tensor $D_{\mu}\tau^{\idxI}D_{\nu}\tau_{\idxI}$; the tensor preserves this form for $\tau^{{\idxI}} \rightarrow \Lambda^{\idxI}_{\ph{\idxI}\idxJ}\tau^{\idxJ}+ S^{\idxI}$ where $\Lambda^{\idxI}_{\ph{\idxI}\idxJ}$ is an orthogonal matrix and $\partial_{\mu}\Lambda^{\idxI}_{\ph{\idxI}\idxJ}=\partial_{\mu}S^{\idxI} =0$. It is tempting to interpret these solutions as a `special-relativistic' limit of the theory in which the dark matter effect is absent.


%

In the symmetry-broken regime $\myt_{\idxI}\myt^{\idxI}<0$ with Friedmann-Lema\^itre-Robertson-Walker symmetry the system~(\ref{grdust}) admits solutions that contain an early time singularity. An important question is the fate of this singularity in the quantum theory, perhaps similarly to how cosmological singularities in systems of gravity coupled to a scalar field may be avoided by quantum geometrical effects~\cite{Ashtekar2003,Singh2009,Achour2014,WilsonEwing2015}. It is additionally conceivable that quantum effects could also address the formation of caustics in the theory, for example producing `$(\square \tau)^{2}$' modifications to the dust Lagrangian such as those considered in~\cite{Chamseddine2014,Chamseddine:2016uef,Brahma:2018dwx}. Furthermore, whereas the metric action~(\ref{grdust}) is not obviously quantizable it might be possible to tackle this problem in the present framework.


The spontaneous symmetry breaking of the complexified Lorentz group is induced in this model by the non-vanishing vacuum expectation value developed by a scalar field that transforms non trivially under the symmetry group, in complete analogy with the Higgs mechanism of particle physics. The parallelism with the latter is however broken by the fact that the predictions of the theory are less directly sensitive to the specific value acquired by the vacuum expectation value in the broken phase. To illustrate this point, consider the effect on the Lagrangian of a field redefinition $\myt^{2}\rightarrow \myt^{2}+\delta {\cal C}$, for constant $\delta {\cal C}$. This can be produced by the variation $\myt^{\idxI}\rightarrow \myt^{\idxI}+\delta\myt^{\idxI}$ where $2\myt^{2} \delta \myt^{\idxI} = \myt^{\idxI}\delta{\cal C}$ and results in a change which is simply a boundary term and does not affect the form of the equations of motion.

In terms of the dynamical ingredients of the theory, one may wonder whether $\myt^{\idxI}$ itself is a composite field. One possibility would be that it could arise from a Dirac spinor $\Psi$ as $\myt^{\idxI} = \bar{\Psi}\gamma^{\idxI}\Psi$; indeed such an object was considered alongside variables $\{e^{\idxI},\omega^{\idxI\idxJ}\}$ as a way of creating a dynamical preferred frame~\cite{Alexander2012}.

\section{Conclusions}
\label{sec:end}

To summarise, we have shown that the spontaneous symmetry breaking of the complexified Lorentz group, in an approach influenced by Cartan geometry, predicts GR plus mimetic DM. This model is related to the Ashtekar formulation of GR via the replacement $e^{\idxI} \leftrightarrow D\myt^{\idxI}$. From a phenomenological point of view, DM arises as a geometric effect which mimics the properties of pressure-less dust, whose abundance is a free parameter of the theory; in a way DM is simply an artifact of how Gravity works. 

Our formulation has the important merit that it makes it possible to investigate the symmetric phase of the theory, where $\myt^{\idxI}=0$ implies that the metric tensor vanishes identically. Indeed, our action is polynomial and there is no explicit inverse metric, so in principle it is possible to smoothly connect the two phases without encountering unphysical singularities.

\section*{Acknowledgements}

We thank Pavel Jirou\v{s}ek, Sabir Ramazanov, and Hans Westman for useful discussions, Carlo Marzo and Iggy Sawicki for critically reading our manuscript, and Alex Vikman for both. LM and TK are supported by ERDF CoE grant TK133. LM is additionally supported by the Estonian Research Council grant PUT 808. FU is supported by the European Regional Development Fund (ESIF/ERDF) and the Czech Ministry of Education, Youth and Sports (MEYS) through Project CoGraDS - CZ.02.1.01/0.0/0.0/15\textunderscore003/0000437. TZ is funded by the European Research Council under the European Union’s Seventh Framework Programme (FP7/2007-2013) / ERC Grant Agreement n.~617656 ``Theories and Models of the Dark Sector: DM, Dark Energy and Gravity''.

\section*{Appendix}

Our conventions for the indices are as follows. We use $\idxI,\idxJ,\idxK$ for indices in the fundamental representation of $SO(1,3)$ (or $SO(1,3)_{C}$); Greek letters are used to denote spacetime coordinate indices; we will use indices $\idxa,\idxb,\idxc$ for spatial coordinates $x^{\idxa}$ in the $3+1$ decomposition, and $\idxi,\idxj,\idxk$ as indices in the fundamental representation of the $SO(3)_{C}$ subgroup of $SO(1,3)_{C}$ that preserves our gauge condition (see main text). Lorentz indices are lowered with the $SO(1,3)$-invariant (as well as $SO(1,3)_{C}$-invariant) matrix $\eta_{\idxI\idxJ}\equiv \mathrm{diag}(-1,1,1,1)$, and antisymmetrisation is defined as $A_{[\idxI}B_{{\idxJ}]}\equiv (A_{\idxI}B_{\idxJ}-A_{\idxJ}B_{\idxI})/2$.

For a differential form $F^{\idxI\idxJ}$ valued in the Lie algebra of $SO(1,3)$ one can decompose the form into \emph{self-dual} (+) and \emph{anti self-dual} (-) parts as $F^{\idxI\idxJ} = F^{+\idxI\idxJ} + F^{-\idxI\idxJ}$, $F^{\pm\idxI\idxJ} =\frac{1}{2}(F^{\idxI\idxJ} \mp i\eps^{\idxI\idxJ}_{\ph{\idxI\idxJ}\idxK\idxL}F^{\idxK\idxL}/2)$, where $ \eps^{\idxI\idxJ}_{\ph{\idxI\idxJ}\idxK\idxL}F^{\pm\idxK\idxL} = \pm 2i F^{\pm\idxI\idxJ}$ define the self-dual and anti self-dual forms. Moreover, we have that $R^{+\idxI\idxJ}_{\ph{+\idxI\idxJ}}(\omega) = R^{\idxI\idxJ}_{\ph{\idxI\idxJ}}(\omega^+)$, and notice that the anti self-dual connection appears in $D = d+ \omega^{+}+\omega^{-}$.  If we had chosen the anti self-dual (-) part of the connection, that is, if we had had $R^{-\idxI\idxJ}_{\ph{-\idxI\idxJ}}(\omega) = R^{\idxI\idxJ}_{\ph{\idxI\idxJ}}(\omega^-)$ in the action, the physics would be unchanged.

\bibliographystyle{hunsrt}
\bibliography{tzreferences}

\end{document}